\documentclass{llncs}

\begin{document}

\title{A Survey of Approaches and Challenges in the Real-time Multimedia Streaming}
%
%
\author{Homayun Motameni \and Mirmorsal Madani
}
%
%

%
\institute{Azad University, Gorgan - Iran\\
}

\maketitle              

\begin{abstract}
We present a brief summary of current approaches and challenges in the network traffic management area. In this contribution, we well cover the most recent experiments on the network traffic and behavior of applications under various network conditions. Quality of the users' experience is the most important factor being considered in this group of experiments.

\end{abstract}
\section*{Introduction}
The amount of sources generating multimedia streaming data has extremely increased in recent decades. Multimedia streaming data are really relevant in computerized systems, social medias, web forums, computational vision applications, and daily digital news purposes to name just a few. Multimedia streaming problems may be approached with intelligent strategies and algorithms along with different network models which required to accurately evaluate and validate. 

Performance evaluation of network applications has received much less attention than applications themselves. Although application developers try to improve the quality of user experience with their products based on the feedback they get from customers. Experimental evaluation of this group of applications needs more attention. Time efficiency and accuracy are two important factors that directly affects the quality of user's experience with network applications \cite{apt5}, \cite{apt1}, \cite{apt2}, \cite{apt3}, \cite{apt4}. 

The motivation of this research study is to revisit the state of the art real-time multimedia streaming systems to provide better knowledge and insights on performance evaluation criteria and address both challenges and possible improvement in the literature. In the current paper, we limit ourselves to perform a survey on experimental evaluations and measurements of only internet-based applications. 
\section*{Related Work}
\par Akhshabi \textit{et al.} performed an experimental evaluation of rate adaption algorithms for streaming over HTTP~\cite{akhshabi2011experimental}~\cite{akhshabi2012experimental}.
The study experimentally evaluates three common video streaming applications under a different bandwidth value ranges. Results of this experiment shows that TCP congestion control and its reliable nature does not necessarily affect the performance of this group of streaming applications. Interaction of rate-adaption logic and TCP congestion control is an area of investigation that has been left for the future.

\par Cicco \textit{et al.} performed an experimental investigation on the Google Congestion Control (GCC) in the RTCWeb IETF WG~\cite{de2013experimental}. In this experiment, authors implemented a controlled testbed. Results of this experimental study show thatGoogle Congestion Control (GCC) in the RTCWeb IETF WG works well but it does not perform a fair bandwidth utilization when the bandwidth is shared by two GCC flows or it is shared by one GCC and one TCP flow.
\par Cicco \textit{et al.} have also performed an experimental investigation on the High Definition (HD) video distribution of Akamai~\cite{de2010experimental}. 

\par Lukasz Budzisz \textit{et al.} proposed a delayed-based congestion control solution for streaming applications~\cite{budzisz2011fair}. Using this system results shorter queues and low delay in homogeneous networks, and balanced flows in delay-based and loss-based heterogeneous networks. Authors of this paper argue that this system can achieve these properties under the whole diversity of loss values, and it outperforms TCP flows. They demonstrate that this system guarantees aforementioned properties using experiments and analysis, .

\par Yang Xu \textit{et al.} performed a measurement study on Google+, Skype, and iChat ~\cite{xu2012video}. In this study, they explained the anatomy of these applications.  The authors of this paper explored some performance details of these applications such as generating the video and techniques used for, strategies to recover from the packet loss, and end-to-end delays metrics Using passive and active experiments.  Based on this experiment, the server location has a significant impact on user performance and also recovery from the packet in server-based applications. This experiment also argues that using batched re-transmissions can be a better solution than Forward Error Correction (FEC) in real time applications. FEC is an error control technique in streaming over unreliable network connections.

\par A mesh-pull-based P2P video streaming solution that uses Fountain codes is presented by Oh \textit{et al.}~\cite{oh2011mesh}. The proposed system offers a fast and smooth streaming application with a low complexity. Evaluations show that the system has better performance than existing buffer-map-based streaming systems when packet loss happens. Considering jitter as another parameter and evaluation of behavior of proposed system under jitter values can be an extension of the study.

\par Performance of Skype's FEC mechanism has been measured by Te-Yuan Huang \textit{et al.} ~\cite{huang2010could}. Authors measured the amount of the redundancy resulting by the FEC mechanism and the trade-offs between the quality of the users' experience and redundancy resulted by the FEC. The study tries to find an optimal level of redundancy to gain the highest quality of experience.

\par Using Fountain Multiple Description Coding (MDC) in video streaming applications over a heterogeneous peer to peer networks is evaluated by Smith \textit{et al.}~\cite{smith2012limit}. This experimental study concludes that Fountain MDC code is a good option in such cases, but there are some restrictions that should be considered in real-world P2P streaming systems. 

\par Assefi \textit{et al.} performed experimental evaluation on real-time cloud speech recognition applications. They used speech recognition applications as a tool for measuring the quality of user experiment under difficult network conditions. The also presented a network coding solution to improve the accuracy and delay of streaming system under different values of packet loss and jitter\cite{assefi2015impact,assefimeasuring,assefi2015experimental}.

\par Te-Yuan Huang \textit{et al.} performed an experimental study on voice rate adaption of Skype under different network conditions~\cite{huang2009tuning}. Results of this experiment shows that using public domain codec is not an perfect solution to achieve user satisfaction. In this study, authors considered different values of packet loss in the experiment and proposed a model to reduce the redundancy resulting from the packet loss reduction.

\par An experimentally study on the performance of multipath TCP over wireless networks is performed by Chen \textit{et al.} ~\cite{chen2013measurement}. They measured the resulting delay from different cellular data providers. Results of this study show that MPTCP results a robust data transport under various network traffic conditions. Studying the  cost of energy and performance trade-offs should be considered as a possible future work for this study. Authors explained details of Akamai's client-server protocol which implements a quality adaption algorithm. This study shows that the proposed technique encodes any video at a number of different bit rates and stores all encodings at the server. Server selects the bitrate matching the bandwidth that is measured based on the receiving signal from the cilent. The bitrate level changes based on the available bandwidth. Authors also evaluated the dynamics of the algorithm in three different scenarios.

\par An algorithm tolerating non-congestion related packet loss is proposed by Hayes \textit{et al.} ~\cite{hayes2010improved}. The experimental evaluation of the algorithm shows that it improves the throughput by 150\% under packet loss of 1\% and allows the system to share the capacity by more than 50\% with respect to other solutions.

A set of experiments to asses quality of experience on television and mobile applications is presented by Winkler \textit{et al.} ~\cite{winkler2003video,winklr2003video}. The proposed experiment considers different values of bit rates, codecs, contents, and network traffic conditions. Authors used Single Stimulus Contiguous Quality Evaluation (SSCQE) and Double Stimulus Impairment Scale (DSIS) on a same set of inputs and compared the results and analyzed methods the compare performance of codecs.

\par A framework for measuring users' QoE is proposed by Kuan-Ta Chen \textit{et al.} ~\cite{chen2009oneclick}. The framework is called OneClick, and  it provides a dedicate key that can be pressed by users every time they feel unsatisfied with the quality of the streaming media.  OneClick is implemented on two applications --shooter games, and instant messaging. 

\par Another setup quantifying the quality of a user's experience is proposed by Kuan-Ta Chen \textit{et al}~\cite{chen2009crowdsourceable}. This system is able to verify inputs from each participant, so it supports crowd-sourcing. Participation is made easy with this setup, and it also generates interval-scale scores. Authors argue that other researchers can use this framework to improve the quality of a users' experience without affecting the results and achieve a higher diversity of users while keeping the cost low. 

\par Finally, Vukobratovic \textit{et al.} proposed a multicast streaming solution that is based on Expanding Window Fountain (EWF) codes in the real-time multicast~\cite{vukobratovic2009scalable}. Using Raptor-like precoding is addressed as a potential improvement in this area. 
\nocite{assefi2012optimizing} 

\section*{Conclusion}
In this study, the general trends in streaming applications was presented along with issues and challenges associated with achieving an efficient and reliable streaming system. Current streaming systems with their respective features were reviewed, related studies in the area were discussed, and open research areas were pointed out. Potential solutions to some of challenges and issues that have been associated with streaming systems, recent advances in the area and some of the most promising experiments, and studies in the field were also presented.
%

\end{document}